# All-Dielectric Meta-optics for High-Efficiency Independent Amplitude and Phase Manipulation

## Author Information

### Affiliations

**Department of Electrical Engineering and Computer Science, University of Michigan, Ann Arbor, MI, 48109 USA**

Brian O. Raeker[*], Anthony Grbic[‡]

**Department of Electrical Engineering and Computer Science, Vanderbilt University, Nashville, TN, 37212 USA**

Hanyu Zheng[*]

**Photonics Initiative, Advanced Science Research Center, City University of New York, New York, NY, 10031, USA**

You Zhou

**Center for Nanophase Materials Sciences, Oak Ridge National Laboratory, Oak Ridge, Tennessee 37831, USA**

Ivan I. Kravchenko

**Department of Mechanical Engineering, Vanderbilt University, Nashville, TN, 37212 USA**

Jason Valentine[†]

### Author Contributions

*these authors contributed equally to this work.
B.O.R. developed device designs, conducted simulations, and wrote the manuscript. H.Z. and Y.Z. performed device fabrication and experimental measurements. I.I.K. performed the material growth. J.V. and A.G. conceived the research concepts and supervised the project. All authors discussed research results and participated in preparing the manuscript.

### Corresponding author

Correspondence to: [‡]Anthony Grbic, agrbic@umich.edu
[†]Jason Valentine, jason.g.valentine@vanderbilt.edu

# Abstract

Metasurfaces, composed of subwavelength scattering elements, have demonstrated remarkable control over the transmitted amplitude, phase, and polarization of light. However, manipulating the amplitude upon transmission has required loss if a single metasurface is used. Here, we describe high-efficiency independent manipulation of the amplitude and phase of a beam using two lossless phase-only metasurfaces separated by a distance. With this configuration, we experimentally demonstrate optical components such as combined beam-forming and splitting devices, as well as those for forming complex-valued, three-dimensional holograms. The compound meta-optic platform provides a promising approach for achieving high performance optical holographic displays and compact optical components, while exhibiting a high overall efficiency.

# Introduction

Controlling the amplitude, phase, and polarization of an optical field is essential to many applications in a broad range of scientific and industrial areas. Conventionally, this is achieved using a sequence of optical components such as lenses, polarizers, gratings, and amplitude masks. Such an approach results in physically large systems as well as reduced efficiency, especially if transmission loss is used to control the amplitude profile of the optical wave. As technological demand increases for compact and high-efficiency optical systems, optical metasurfaces offer a path toward satisfying such requirements. Metasurfaces are two-dimensional arrays of subwavelength scatterers, designed to exhibit remarkable control over the transmitted phase, amplitude, and polarization of light. Specifically, the ability to locally control the transmission characteristics provides a powerful approach to realizing flat optical components and systems with improved performance[1–4]. This has been shown through a variety of devices, including flat lenses[5,6], beam-splitters[7–12], holograms[13–17], augmented reality displays[18], high-definition displays[19], image differentiation[20], and compact optical spectrometers[21]. However, optical applications requiring spatial manipulation of the amplitude profile have relied on reflection[22], absorption[23], or polarization conversion loss[11,12,14–17], resulting in low efficiencies.

Independently controlling the spatial amplitude and phase distributions of an electromagnetic wave is essential to forming optical components such as combined beam-former and splitter devices. Metasurface beam-splitters have been demonstrated through a variety of methods[7–12], however the beam shape is not altered, and efficiency is reduced due to diffraction or the use of loss. Another use of amplitude and phase control is in forming three-dimensional holograms, which can be produced with high quality if a specific complex-valued field profile is formed[16]. Polarization loss has been used to form three-dimensional holograms[14–17] by performing a spatially varying polarization conversion across the metasurface. Amplitude control is implemented by filtering the unconverted polarization component, while phase control is implemented via another metasurface parameter. While capable of producing high-quality holographic images, the efficiency of such devices is low and application-specific due to reliance on polarization loss.

In this work, we experimentally demonstrate high-efficiency, independent amplitude and phase control over an optical field using compound meta-optics. Compound meta-optics are sequential metasurfaces separated by a distance and arranged along a common axis, which perform functions beyond the capabilities of individual metasurfaces[24,25]. The configuration of the compound meta-optic is shown in Figure 1(a). A process for designing paired phase-only metasurfaces was previously developed to implement independent amplitude and phase control[24,25]. This process was verified through full-wave simulation at near-infrared wavelengths to form high-efficiency combined beam-former and splitters and a high-quality, three-dimensional hologram[25]. Using this approach, the spatial separation of the metasurfaces enables high-efficiency amplitude control by redistributing the optical beam amplitude instead of using loss to form the desired amplitude distribution. Alternative methods have also been developed to optimize sequential metasurfaces for spatial complex-valued control over a wave. Optimizing metasurface unit cell parameters using the adjoint optimization method[26], optimizing the plane wave spectrum[27], and optimizing equivalent current distributions[28], have been used to design sequential metasurface systems. Additionally, a variety of applications have been demonstrated with sequential metasurface devices: aberration correction[29], optical retro-reflection[30], full-color holography[31], and optical diffractive neural networks[32,33].

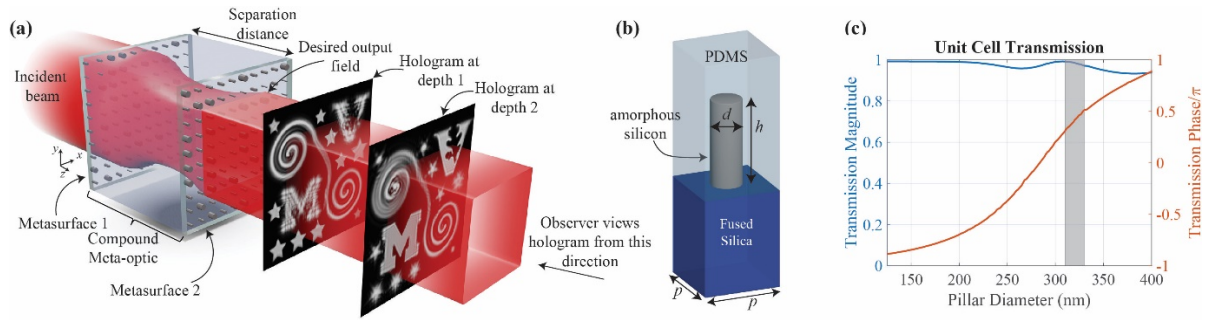

Figure 1: Diagram of the optical compound meta-optic implementing independent phase and amplitude control over an incident optical field. **(a)** Illustration of a compound meta-optic converting a uniform illumination into a three-dimensional, complex-valued hologram. Depth of the three-dimensional hologram is visualized when the output field is imaged at different depths by an observer facing the device. **(b)** A schematic of the metasurface unit cell. Each cell has a period $p = 570nm$, and contains an amorphous silicon nanopillar with a height $h = 850nm$. The nanopillar diameter is varied across the metasurface to implement the desired transmission phase profile. **(c)** The transmission characteristics of the unit cell under normally incident plane wave illumination and periodic boundary conditions. A phase coverage of 78% with transmission above 0.93 is achieved by varying the pillar diameter. Diameters of 310nm-330nm are not used in the three-dimensional hologram designs due to observed low transmission in fabrication, which impact the image quality.

The compound meta-optic is formed by two lossless phase-only metasurfaces separated by a physically short distance of homogeneous dielectric, as shown in Figure 1(a). Together, they provide the desired independent amplitude and phase control. Adding additional metasurfaces would allow more control over the optical field (e.g. multi-wavelength performance[6,22,26], or diffractive neural networks[26,32,33] for multi-input multi-output applications), but only two metasurfaces are necessary for amplitude and phase control at a single wavelength. As a proof of concept, we report experimental demonstrations of meta-optics that combine beam-forming and splitting, and produce high-quality, three-dimensional holograms. The efficiency of the devices is high, only limited by the minor reflection from each metasurface. Here, we show the promise of optical compound meta-optics for high-efficiency holographic displays and optical components.

## Results

**Concept and design of compound meta-optics.**

The proposed compound meta-optic devices consist of two phase-only metasurfaces, each impressing a phase discontinuity onto an incident wave. The design method has been developed and verified in simulation at microwave frequencies[24] and near-infrared wavelengths[25]. The first metasurface applies a phase discontinuity onto the known incident wave. This phase shift is designed so that the desired electric field amplitude distribution is formed at a distance equal to the separation distance between the metasurfaces. However, the phase of this field profile is incorrect relative to the desired output phase profile. Therefore, the second metasurface applies a phase correction to form the desired complex-valued electric field distribution. Since the metasurfaces are assumed to be transparent, the amplitude distributions of the wave transmitted by metasurface 1 and incident on metasurface 2 are known from the incident and desired field profiles, respectively. The free parameters are the transmission phase distributions at each plane, which are optimized using a phase-retrieval algorithm that links the incident and output field amplitude distributions. Specifically, the Gerchberg-Saxton phase-retrieval algorithm[34] is modified to operate over short distances instead of the near-to-far-field conversion typically used. The phase shift distribution of each metasurface is calculated as the difference between the phase distributions of the transmitted and incident fields.

$$\phi_{ms1} = \phi_{tr1} - \phi_{E_{inc}} \qquad (1)$$

$$\phi_{ms2} = \phi_{E_{out}} - \phi_{inc2} \qquad (2)$$

where $\phi_{E_{inc}}$ is the phase of the incident field, $\phi_{tr1}$ is the phase profile of the transmitted field from metasurface 1, $\phi_{inc2}$ is the phase of the field incident on metasurface 2, and $\phi_{E_{out}}$ is the phase profile of the desired output field. The metasurfaces are then designed to implement $\phi_{ms1}$ and $\phi_{ms2}$ as a transmission phase shift of the form $e^{i\phi_{ms}}$ on the incident wave. Here, a time convention of $e^{-i\omega t}$ is assumed.

Each metasurface is a dense array of silicon nanopillars, which are commonly used to implement desired spatial phase and polarization modulations with high transmission efficiency[1–4]. Figure 1(b) shows a schematic of the unit cell, with the silicon nanopillar embedded in a layer of polydimethylsiloxane (PDMS) at the interface of a fused silica handle wafer. Circular cross-sections of the nanopillars were chosen to ensure polarization invariance, however elliptical[35] or rectangular[12,16,22,36] cross-sections could be used if polarization control is required. Variation of the pillar diameter ($d$) provides control over the transmission phase. The silicon pillar height ($h = 850nm$), unit cell period ($p = 570nm$), and illuminating wavelength ($\lambda_0 = 1.3\mu m$) have been chosen to provide high transmittance for a large transmission phase range. The locally periodic approximation[37,38] is made to determine the unit cell transmission by assuming that the nanopillars are placed in a homogeneous array environment. This approximation is commonly used to simplify the design of unit cells in inhomogeneous arrays and is generally accurate if the unit cells are not strongly coupled (as is the case here). Figure 1(c) shows the transmission for the chosen dimensions as a function of pillar diameter. For these dimensions, a transmission phase coverage of 78% can be achieve with a transmission magnitude greater than 0.93. Note that the pillar diameters of $125nm - 400nm$ are used in the beam-splitter designs, but $310nm - 330nm$ are not used in the three-dimensional hologram design. This is due to observed low transmission for these diameters in measurement, which impacts image quality (additional details can be found in Supplementary Note 1).

The phase shift profiles producing the desired amplitude and phase conversion of the input field are sampled at the unit cell periodicity and converted to distributions of nanopillars with the corresponding diameters. Each metasurface is then fabricated individually and aligned to form the compound meta-optic devices.

**Fabrication and characterization of compound meta-optics**

The meta-optic devices were fabricated using nanofabrication techniques developed to construct multi-metasurface devices[6,22]. First, an 850nm-thick layer of amorphous silicon was deposited onto a fused silica substrate. Each metasurface pattern was defined using electron beam lithography (EBL) and then nanopillars were formed using reactive ion etching (RIE)[6,20,22]. The metasurfaces were subsequently enclosed in a protective layer of PDMS. Figure 2(a) shows optical images of the meta-optic devices and one fabricated metasurface, and a scanning electron microscope image of the circular silicon nanopillars after RIE. Finally, the metasurfaces were spaced by a layer of PDMS and carefully aligned using translation stages to form the complete compound meta-optic, as shown in Figure 2(b). Accurate alignment between the two metasurfaces ($< 1.5\mu m$ lateral misalignment) is necessary to obtain the desired complex-valued field transformation. To achieve this, we developed hologram alignment marks formed by silicon arrays fabricated near each of the two metasurfaces (additional information can be found in Supplementary Note 2). After spatial alignment was achieved with the alignment marks, the alignment was further adjusted until the desired intensity image was formed at a distance beyond the meta-optic output. Misalignment introduces phase error to the output field distribution and the output field phase defines the propagation behavior of the wave, so the alignment improves as the observed intensity image improves.

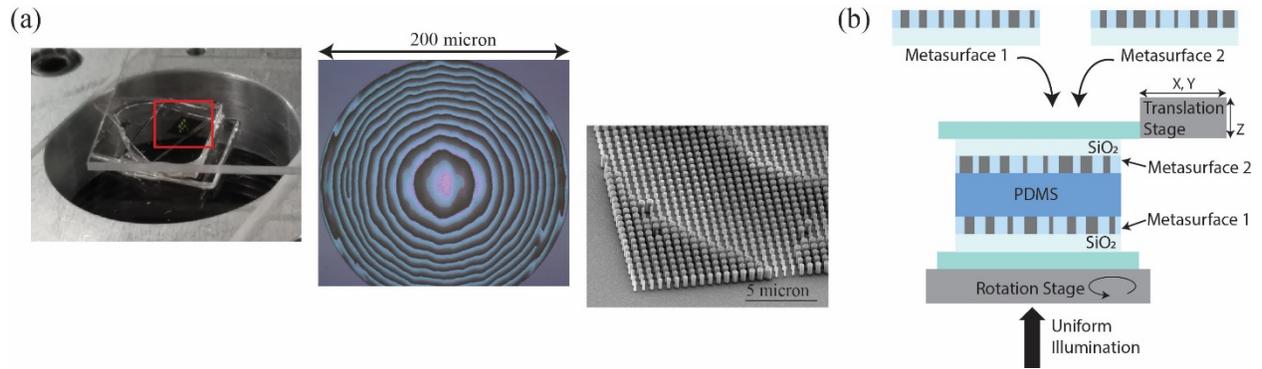

Figure 2: Images of a fabricated meta-optic and schematic of meta-optic assembly. **(a)** Left: An optical image of the fabricated meta-optic device, where the two metasurface layers are being aligned. Middle: Optical image of a fabricated metasurface, showing the variation in pillar diameter as a change in color. Right: A scanning electron microscope image of a portion of one metasurface, showing the array of silicon nanopillars. **(b)** Schematic of the meta-optic assembly, alignment, and characterization process. Each metasurface is fabricated individually and then aligned to form the meta-optic device.

The compound meta-optics were characterized using an unpolarized supercontinuum laser whose beam is passed through a monochromator to select the desired wavelength. The resulting beam overfilled the meta-optic footprint to form the desired uniform illumination. The meta-optic performs the required spatial amplitude and phase manipulation to form the desired complex-valued output field distribution. Intensity distributions at different depths from the meta-optic output were then magnified with an objective and tube lens and recorded with a camera. Amplitude control over the incident optical field can be directly verified by imaging the output plane of the meta-optic. If the output field phase distribution is accurate, propagation of the output field will form the desired intensity profile at each plane in space beyond the meta-optic. However, if the phase distribution is inaccurate, the desired intensity will not be formed at a distance beyond the meta-optic. Therefore, phase control was verified by comparing the measured intensity distribution at a plane beyond the meta-optic output.

To demonstrate the accuracy of the compound meta-optic in providing independent phase and amplitude control with high efficiency, two different functions are presented. First, we show a combined beam-forming and splitting function where a uniform illumination is reshaped to form multiple output beams with specified amplitude profiles and propagation directions. Second, we design meta-optics to reshape a uniform illumination to form computer-generated, three-dimensional holograms.

**Meta-optics for combined beam-forming and splitting**

Using the meta-optic design process[24,25], we designed, fabricated, and measured two meta-optic devices performing optical beam-forming and splitting at a wavelength of $\lambda_0 = 1.3\mu m$. In both examples, a circular uniform illumination with diameter of $200\mu m$ is reshaped to form multiple output Gaussian beams. Each metasurface is $200\mu m$ in diameter and separated by $325\mu m$ of PDMS. The first meta-optic forms the interference pattern between two Gaussian beams of different beamwidths, propagation directions, and relative intensities. The desired output field profile is calculated as the superposition of the two Gaussian beams as

$$E_{out} = -e^{-(r/45.5\lambda_0)^2}\,e^{ik_0 x\sin(-2°)} + \sqrt{0.5}e^{-(r/32.5\lambda_0)^2}\,e^{ik_0 x\sin(2°)} \qquad (3)$$

which forms a fringed interference pattern. Since the exact complex-valued field distribution is formed, only the desired Gaussian beams are produced, and no undesired diffraction orders are present. Each metasurface of the device was simulated using the open-source finite-difference time-domain (FDTD) solver MEEP[39], and their responses combined to determine the overall meta-optic output field distribution (additional information about the meta-optic designs can be found in Supplementary Note 3). Figure 3(a) compares the intensity profiles formed by the meta-optic at the output plane and at a far-field distance for simulation and measurement. In each case, we see that the measurements closely agree with the simulated intensity distributions and form the desired Gaussian beams with little diffractive noise. Furthermore, the transformation in amplitude and phase was performed with an efficiency of 81% in simulation and 72% in measurement. The efficiency is defined as the fraction of intensity in the uniform illumination contained within the beam cross-sections (with Fresnel reflection correction).

A second beam-former/splitter example was fabricated to produce three output Gaussian beams. The propagation directions are chosen such that the output field has variation in both planar dimensions (see Supplementary Note 3 for design specifics). Figure 3(b) compares the intensity profiles formed by the meta-optic at the output plane and a far-field distance in simulation to measurements. Here, the desired Gaussian beams are formed at the correct relative intensities, propagation directions, and beamwidths, verifying the accuracy of amplitude and phase control performed by the meta-optic. The device efficiency is 79% in simulation and 65% in measurement. One reason for the slight reduction in experimental efficiency of the beam-splitter devices is an experimentally observed transmission dip for pillars with diameters in the $310nm - 330nm$ range. This feature is discussed in more detail in Supplementary Note 1.

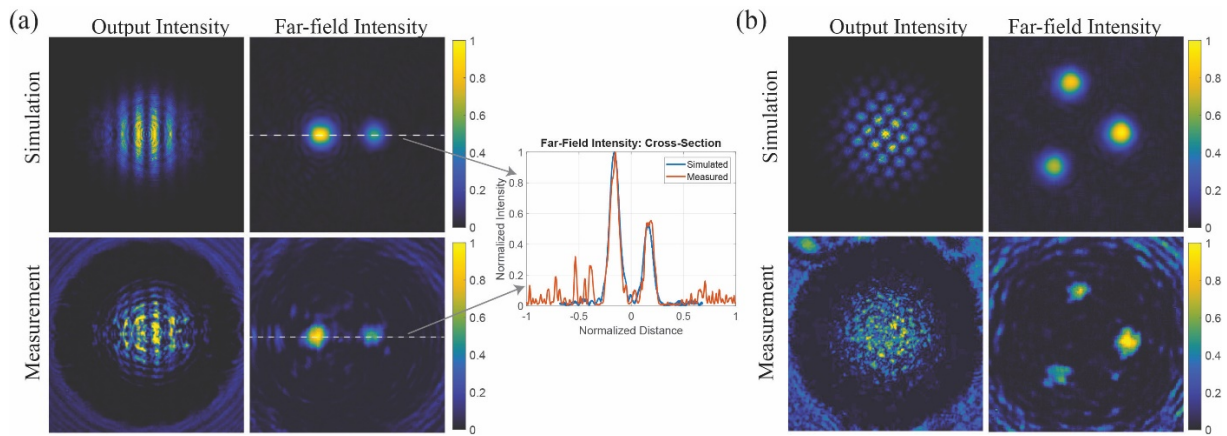

Figure 3: Measurement results of meta-optics implementing a beam-forming and splitting function, where a uniform circular illumination is manipulated to form multiple Gaussian beams at the output. In each case, measurements closely match the simulation results. Note that intensities have been normalized to respective maximums. **(a)** Two Gaussian beams are formed by the meta-optic. The output intensity is the characteristic fringing pattern of two interfering beams, verifying that the amplitude distribution of the optical field has been reshaped. The two beams appear separate at a far-field distance, verifying that the phase distribution at the output is also correct. A cross-section through the beams is taken to show that the Gaussian profile of each beam is accurately formed. **(b)** Three Gaussian beams are formed by a meta-optic. The output intensity is the interference pattern between the three beams and verifies that the amplitude profile has been reshaped. Three separate beams are observed at a far-field distance, matching simulation results. (Perceptually uniform color bar[40] used for these and all following plots).

**Meta-optics for three-dimensional holograms**

Another exciting application of high-efficiency amplitude and phase control is three-dimensional holography. While phase-only metasurfaces can produce holograms, phase and amplitude control allows for enhanced image quality, especially for three-dimensional holograms[13,16]. Such complex-valued holograms have been generated using lossy methods, most commonly reflection[22] or polarization loss[14–17]. Here, we utilize the meta-optic design approach to demonstrate high-efficiency, three-dimensional holograms. Multiple computer-generated hologram approaches can be used to display a three-dimensional scene[41], but we demonstrate two methods: a simple point-source hologram and a hologram composed of solid flat image components[42–44] more applicable to generating a realistic hologram scene.

Point-source holograms approximate the surface of a three-dimensional object with a collection of point sources and have the advantage of wide viewing angles due to the diverging spherical wave from each point source. The field distribution at the meta-optic output that forms the hologram is calculated by summing the complex conjugated radiated fields from each point source. Phase-only metasurfaces can implement such holograms by directly applying the phase of this field distribution to the incident field to form images. However, image quality is notably improved if the amplitude pattern is also formed.

Here, a compound meta-optic is designed to form the complex-valued interference field of point sources tracing the outline of the University of Michigan and Vanderbilt University logos. Each logo is tilted about its center by 15 degrees to provide depth to the image, as shown by the diagram in Figure 4(a). The metasurfaces, designed to operate at $\lambda_0 = 1.1 \mu m$, were square arrays $200 \mu m$ in extent and separated by a distance of $275 \mu m$ (see Supplementary Note 3 for additional design details and different unit cell dimensions). Figure 4(b) compares the simulated hologram to the measured hologram at different depths, where the in-focus portion of each image is marked with a red arrow. We see that the measured intensity images closely match the desired intensity images with very little speckle noise.

While point-source holograms have wide viewing angles, they are unable to visually re-create the appearance of a realistic scene due to the point-based sampling of the surface. Next, we use computer-generated hologram techniques to generate a three-dimensional hologram consisting of solid image components tilted in space[42–44]. These techniques can be used to produce a faceted representation of an object, leading to the ability of forming 3D holograms of life-like scenes[25,42–44].

For this example, we designed a compound meta-optic where each metasurface is a $200 \mu m$ square array of nanopillars and separated by $325 \mu m$ of PDMS. The operating wavelength is $\lambda_0 = 1.3 \mu m$. The output field of the compound meta-optic was engineered in amplitude and phase to form the scene shown in Figure 4(c). The large spiral, University of Michigan (M), and Vanderbilt University (V) logos are all tilted in space, but in different directions (see Supplementary Note 3 for additional design details). As a result, different cross-sections of the image come into focus when imaging different depths from the meta-optic output plane. Figure 4(d) shows intensity images measured at different depths in the 3D hologram, which closely match the simulated images. Specifically, the first image shows the middle of the spiral and stars in focus, while the remaining images show different cross-sections of the logos. The full three-dimensional nature of the hologram can be seen by scanning the focal plane through the volume of the hologram (see Supplementary Movie 1). The compound meta-optic performs the desired complex-valued field manipulation with a simulated efficiency of 82% and measured efficiency of 75% for this example, significantly exceeding the efficiency of loss-based approaches[14–16,22]. The efficiency considered here is the percentage of the input intensity present in the transmitted field distribution at the output of the meta-optic.

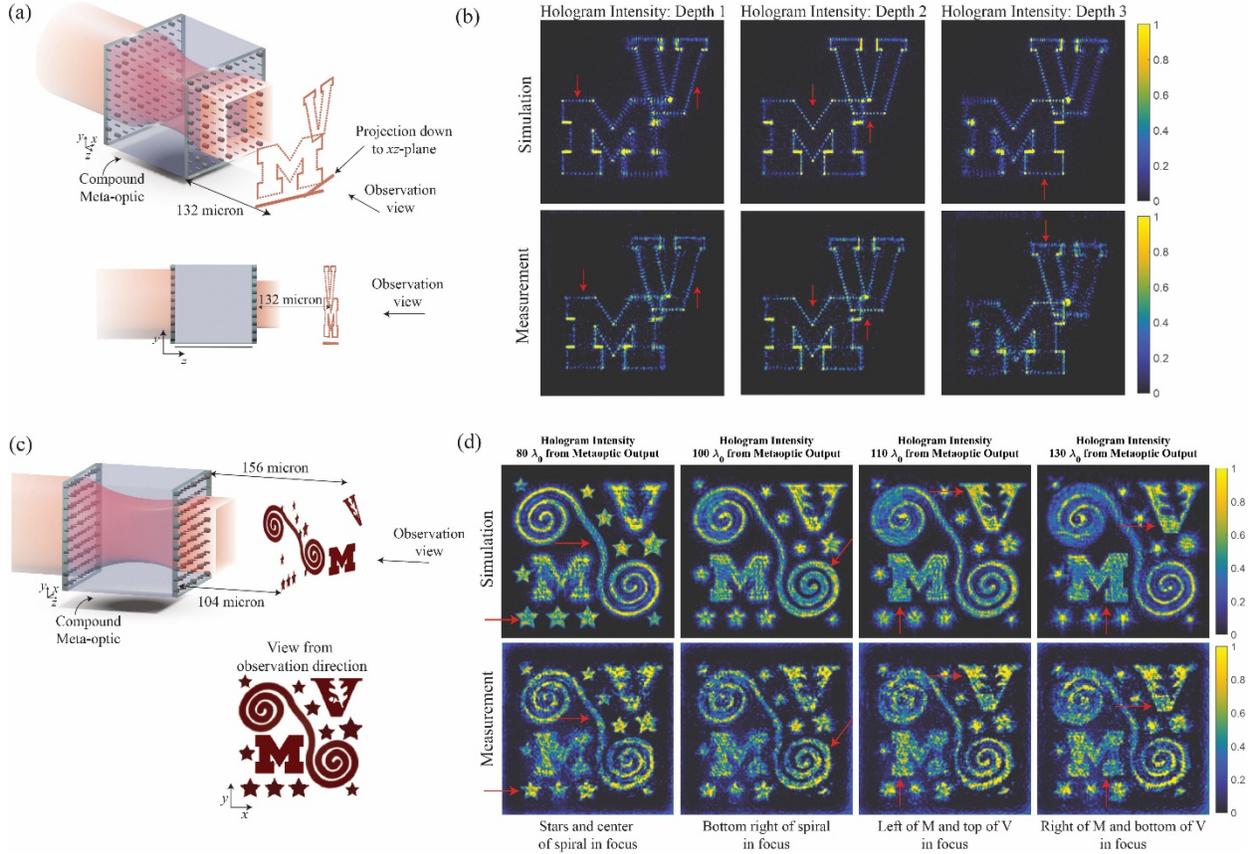

Figure 4: Comparison of simulation and measurement results of compound meta-optics implementing two 3D holograms. In each case, the measurement results closely match the expected simulation results. Note that intensities are normalized to respective maximums. **(a)** A point source hologram formed by a compound meta-optic at $\lambda_0 = 1.1\mu m$. The University of Michigan and Vanderbilt University logos are tilted in space by 15 degrees, providing depth to the hologram. **(b)** The hologram intensity was recorded at different distances from the output plane of the compound meta-optic, showing that different portions of the hologram are in-focus at different depths (as denoted by the red arrows). **(c)** A 3D hologram using solid image components is formed by a compound meta-optic at $\lambda_0 = 1.3\mu m$. The large spiral, University of Michigan logo, and Vanderbilt University logo are tilted in space to provide depth to the scene, while the stars are formed in the same plane, at the halfway point of the spiral. **(d)** The hologram intensity was recorded at different distances from the meta-optic output, showing the different image components coming into and out of focus.

## Discussion

The measured intensity images from each of the meta-optic devices demonstrate the accuracy of complex-valued optical field control with high efficiency. Specifically, the intensity measured at the output plane of the meta-optic demonstrates amplitude control since the uniform illumination is reshaped. The intensity measured at a plane beyond the aperture (after the optical field propagates away from the meta-optic output) demonstrates phase control since the desired intensity will only be formed if the output phase profile is accurate. An inaccurate phase profile would alter the propagation characteristics of the optical field so that the measured intensity pattern would not match the desired result. This is verified in Figure 3, where combined beam-former and splitters are demonstrated to form intensity distributions which closely match the expected distributions. Even though a small number of Gaussian beams were formed with these devices, meta-optics can be designed to produce a larger number of beams with different amplitude profiles[25]. The measured efficiency of the beam-former/splitter designs of 78% and 65% is higher than that of methods which would require loss.

Similarly, the measured three-dimensional holograms shown in Figure 4 very closely match the desired intensity images, demonstrating that the meta-optics can produce life-like holograms. The images produced in Figure 4(d) with phase and amplitude control avoid the image speckle common to phase-only implementations produced with phase-retrieval algorithms[16]. Furthermore, scanning the imaging plane through the hologram shows the image components are accurately reproduced at the desired spatial locations (see Supplementary Movie 1). This approach also maintains the image contrast compared to the desired phase profile simply applied to the uniform illumination (see Supplementary Note 4 for comparison). While the fabricated meta-optics demonstrated relatively simple holograms, meta-optics can be designed to form 3D holograms of life-like scenes[25].

High-efficiency amplitude and phase control over optical fields enables compound meta-optics to explicitly perform a variety of applications without inherent losses in efficiency. In contrast, loss-based amplitude control significantly reduces the device efficiency. For example, a measured efficiency of 6.4% was demonstrated in forming a complex-valued hologram using polarization conversion at THz frequencies[17] (however, intrinsic unit cell losses also contributed, since the measured efficiency for a similar phase-only version was 19.1%). Even when unit cells are optimized to obtain a maximum polarization conversion efficiency of 100%[16], the overall efficiency of a device is less than unity and highly case-dependent due to variation of transmission amplitude over the metasurface. Therefore, there are two main advantages of the compound meta-optic approach in terms of efficiency. First, using polarization loss for amplitude control requires a specific input polarization, which decreases the overall efficiency if an unpolarized source is used. Instead, the method shown here is polarization independent and avoids this issue. Second, the efficiency of loss-based approaches is highly case-dependent, but compound meta-optics can achieve near-unity efficiency (in the ideal case) regardless of the required amplitude control. As a result, the measured 75% efficiency of the compound meta-optic producing the solid-image 3D hologram is significantly higher than lossy methods[14–17] and is approximately independent of the desired amplitude control.

The reduced measured efficiency relative to simulation is attributed to fabrication errors, in particular an experimental transmission dip for pillars with diameters in the $310nm - 330nm$ range (in the case of the beam-splitting devices), and misalignment of the two metasurfaces. Fabrication errors and misalignment can also lead to phase errors in the output field distribution, causing slight degradation in image quality observed between simulation and experiment. Additionally, since a locally periodic approximation was used to design the metasurfaces, differences between the local periodic performance of each unit cell and inhomogeneous metasurface performance contribute to reduced efficiency and image quality reduction. Optimizing the inhomogeneous metasurface structure[45,46] could lead to improvements in efficiency and transmission phase accuracy. That said, the demonstrated compound meta-optics already exhibits amplitude and phase control with high efficiency and accuracy.

In conclusion, we have demonstrated compound meta-optics consisting of paired, lossless metasurfaces that independently manipulate the amplitude and phase distributions of an optical field. Each metasurface was implemented as a high-transmission array of amorphous silicon nanopillars and aligned, operating at near-infrared wavelengths. The distance between the metasurfaces allows the optical wave to be reshaped, leading to high-efficiency devices by avoiding loss-based amplitude-control mechanisms. High-efficiency field control expands the application space of meta-optics, while maintaining a compact form-factor. As examples, we have experimentally shown that compound meta-optics can implement optical functions such as combined beam-forming and splitting, as well as form three-dimensional holograms with high image quality. In each case, the measured efficiency of the fabricated meta-optic devices ranged between 65%-75%. By explicitly forming the desired complex-valued field profiles, no energy is lost to diffraction and high-quality holograms are obtained. The compound meta-optic approach can be extended to visible optical wavelengths by a change in materials. For example, titanium dioxide can be used in place of silicon to fabricate the nanopillars[5,47]. Additionally, polarization control can be achieved through an anisotropic nanopillar cross-section. Compound meta-optics could lead to improved performance in three-dimensional holography, compact holographic displays, custom optical elements, and other applications requiring detailed control over the phase, amplitude, and polarization distributions of an optical field.

## Methods

### Simulation

An open-source finite difference time domain (FDTD) solver MEEP[39] was used to simulate the transmission performance of the silicon nanopillar unit cells and each metasurface composing the meta-optics. The unit cells were individually simulated with periodic boundary conditions, while the metasurface simulation used perfectly matched layer (PML) boundaries $2\mu m$ from the metasurface edge. The source field distribution was placed $1.5\mu m$ in front of the metasurface or unit cell, and a monitor placed $1.5\mu m$ after to record the transmission field amplitude and phase. To simulate the compound meta-optic, the transmitted field from the simulation of metasurface 1 was numerically propagated across the separation distance and used as the source for the simulation of metasurface 2[25]. The transmitted field from the FDTD simulation of metasurface 2 is considered the output field of the meta-optic and compared to the measurement results.

## Fabrication

The metasurface patterns were defined on an 850nm-thick amorphous Si wafer, which was grown on a fused silica substrate using low-pressure chemical vapor deposition (LPCVD). A 200nm-thick PMMA A4 was spin coated at 4500 rpm, followed by deposition of 10nm-thick chromium as the conduction layer using thermal evaporation. The metasurface structures were defined using electron beam lithography (EBL), followed by the deposition of 35nm-thick aluminum oxide as the dry etch mask using e-beam evaporation. The Si nanoposts were then formed by reactive ion etching (RIE) using a mixture of $SF_6$ and $C_4F_8$.

To completely encapsulate the Si nanoposts, the first layer of PMDS (10:1 mixing ratio of Sylgard 184 base and curing agent) was diluted in toluene in a 2:3 weight ratio, which was spin coated at 2500 rpm and cured at 80 C° for more than half an hour. The second layer of PMDS without dilution was subsequently spin coated and cured at 80C° for more than 2 hours. The same procedure was used for both metasurface layers.

## Measurement

A customized alignment system is shown in Figure 2b. The bottom metasurface layer was mounted on a rotation stage held by a vacuum pump, and the top layer was attached to a glass slide suspended by an XYZ translation stage. A drop of uncured PMDS was applied in between as an index-matched layer. The samples were then illuminated from the bottom by a collimated supercontinuum laser that was passed through a monochromator set at the designed working wavelength. The far-field images after the bilayer metasurface were recorded by a NIR camera through an imaging system consisting of a 20X objective and a tube lens. The alignment holograms were then aligned by tuning the XYZ translation and rotation stage. After spatial alignment was achieved, the XY translation was further adjusted until the desirable far-field images were formed. The efficiency was calculated based on the images captured by the NIR camera with the background noise correction processed by subtraction of a blank image, which contained the dark current signals. The reference intensity was calculated from images of the substrates without any meta-optics. With the same integration time, the intensity distribution can be revealed by the photon count from the camera. The efficiency of the beam-former/splitter devices was calculated as the intensity within an area encompassing each beam in the far-field, divided by the intensity incident on the device. For the three-dimensional hologram example, the efficiency is the intensity contained in the transmitted field divided by the intensity incident on the device.

## Acknowledgements

B.O.R. and A.G. acknowledge financial support by the National Science Foundation Graduate Research
Fellowship Program. B.O.R, H.Z., Y.Z., J.V., and A.G. acknowledge financial support by the Office of
Naval Research under Grant No. N00014-18-1-2536. Additionally, this research was supported in part
through computational resources and services provided by Advanced Research Computing at the
University of Michigan, Ann Arbor. A portion of this research was conducted at the Center for
Nanophase Materials Sciences, which is a DOE Office of Science User Facility. Finally, B.O.R. and A.G.
are grateful to Cody Scarborough for assistance in creating graphics seen here.

## Additional information

## Competing Interests

The authors declare no competing interests.

# Supplementary Information for: All-Dielectric Meta-optics for High-Efficiency Independent Amplitude and Phase Manipulation

*Brian O. Raeker[1]\*, Hanyu Zheng[2]\*, You Zhou[3], Ivan I. Kravchenko[4], Jason Valentine[5]†, Anthony Grbic[1]‡*

[1]Department of Electrical Engineering and Computer Science, University of Michigan, Ann Arbor, MI, USA

[2]Department of Electrical Engineering and Computer Science, Vanderbilt University, Nashville, TN, USA

[3]Photonics Initiative, Advanced Science Research Center, City University of New York, New York, NY, 10031, USA

[4]Center for Nanophase Materials Sciences, Oak Ridge National Laboratory, Oak Ridge, Tennessee 37831, USA

[5]Department of Mechanical Engineering, Vanderbilt University, Nashville, TN, USA

*these authors contributed equally to this work.

†email: jason.g.valentine@vanderbilt.edu

‡email: agrbic@umich.edu

## Supplementary Note 1: Unit Cell Design and Performance

### S1.1. Simulation of Unit Cell vs. Dimensions

Each compound meta-optic is composed of two metasurfaces applying a polarization-independent phase profile onto the incident wave. The metasurfaces are implemented as arrays of high dielectric contrast nanopillars, which locally provide the desired transmission phase shift by varying the dimensions of the nanopillars across the metasurface. Figure S1(a) shows a schematic of the metasurface unit cell and the relevant dimensions. The unit cell dimensions can be varied to achieve a desired transmission profile as a function of pillar diameter. In this case, a large phase coverage with high transmission magnitude is desired. To find the desired pillar dimensions, simulations of the metasurface unit cell (with Floquet boundary conditions to simulate a periodic array) were performed to determine the transmission magnitude as a function of the pillar diameter and the period, pillar height, or wavelength (see Figure S1(b)-(d)). The transmission phase will change as the pillar diameter changes, so dimensions are sought which maximize transmission magnitude.

Figure S1(b)-(d) show that a high transmission magnitude as a function of diameter can be achieved for a unit cell period of $p = 570nm$, pillar height of $h = 850nm$, and operating wavelength of $\lambda_0 = 1.3\mu m$. The transmission magnitude and phase are shown in Figure S1(e) for these dimensions. By using pillars with diameters ranging from $125nm$ to $400nm$, a transmission phase coverage of 78% with transmission above 0.93 is achieved. Pillar diameters of 310nm-330nm were skipped in the three-dimensional hologram meta-optic design because low-transmission contours were observed in measurement at these diameters. See Section S1.2 for additional details.

The locally period approximation[1,2] was used to simplify the design of the metasurfaces. The pillar diameters of the array were determined by matching the desired local transmission to the transmission obtained from periodic simulations of each unit cell. This approximation is commonly used, with resulting metasurface performance largely matching the desired transmission characteristics. The most significant exception is the contour between the smallest-diameter pillars and the largest-diameter pillars, where simulations of the full metasurface show slightly reduced transmission compared to the local-periodic expectation. However, this is a small error in the transmitted field and does not significantly alter the resulting output field of the meta-optic.

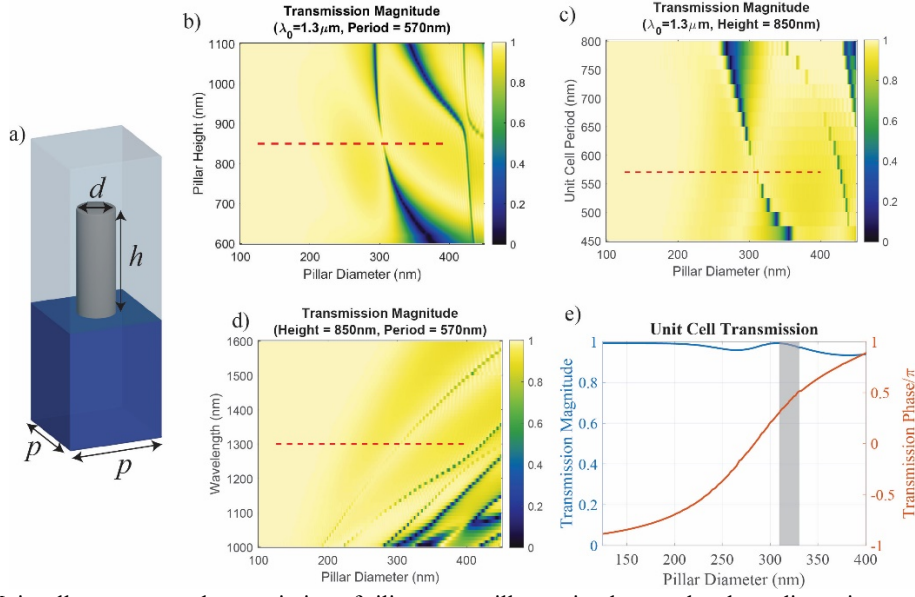

Figure S1: Unit cell geometry and transmission of silicon nanopillars to implement the phase-discontinuous metasurfaces. **(a)** Diagram of the metasurface unit cell, consisting of an amorphous silicon nanopillar at the boundary of a fused silica and PDMS material interface. The pillars have a height of $850nm$ and a spacing of $570nm$. **(b-d)** Plots showing the transmission magnitude of the unit cell for variations in dimension around the nominal design, which is marked with a dashed line. **(e)** Transmission characteristics for the nominal unit cell dimensions while varying the pillar diameter. Diameters of $310nm -$ $330nm$ are avoided in the hologram design since low transmission was observed in fabricated metasurfaces for these diameters. (Perceptually uniform color bar[3] used for these and all following plots).

## S1.2. Measurement of Low-Transmission Contours

Pillar diameters of $310nm$ through $330nm$ are not used in the metasurface design for the solid image hologram meta-optic (example shown in Figure 4(c-d) of the main text). This is due to low transmission intensity measured over contours on fabricated metasurfaces containing this diameter range, which is lower than the expected full transmission in simulation. A low measured transmission magnitude is an indicator that the phase of the transmitted field also deviates from the expected value. Both errors in the measured transmitted field combine to adversely impact the measured output field of the meta-optic.

This effect was observed in a meta-optic fabricated for a hologram design (different design than those presented in the main text) using the entire $125nm - 400nm$ pillar diameter range. The diameters of the nanopillars constituting metasurface 1 are shown in Figure S2(a). The measured transmitted intensity of this metasurface is shown in Figure S2(b), where low transmission contours are visible. Various contours of different pillar diameter ranges were compared to the measured transmitted intensity to determine which pillars were exhibiting low transmission in measurement. Figure S2(c) shows the contour formed by pillars with diameters of 310nm-330nm overlayed with the transmitted intensity. The nanopillars with diameters of 310nm-330nm align with the low-transmission portions of the metasurface, suggesting that the low transmission occurs for these specific nanopillar diameters.

While the cause of low transmission for these pillar diameters is not explicitly known, it is expected that fabrication errors alter the symmetry of the pillar, which disrupts the resonance behavior to introduce a reduced transmission. For example, it has been shown that asymmetry in high-index contrast structures enables coupling to symmetry-protected modes (also known as bound states in the continuum) to produce sharp transmission or reflection resonances[4–6]. The simulations assumed perfectly shaped cylinders, so might not have experienced this effect. However, Figure S1(b-c) show that a slight change in pillar height or pillar spacing can cause low transmission for this range of pillar diameters, so it is expected other fabrication errors could have a similar effect.

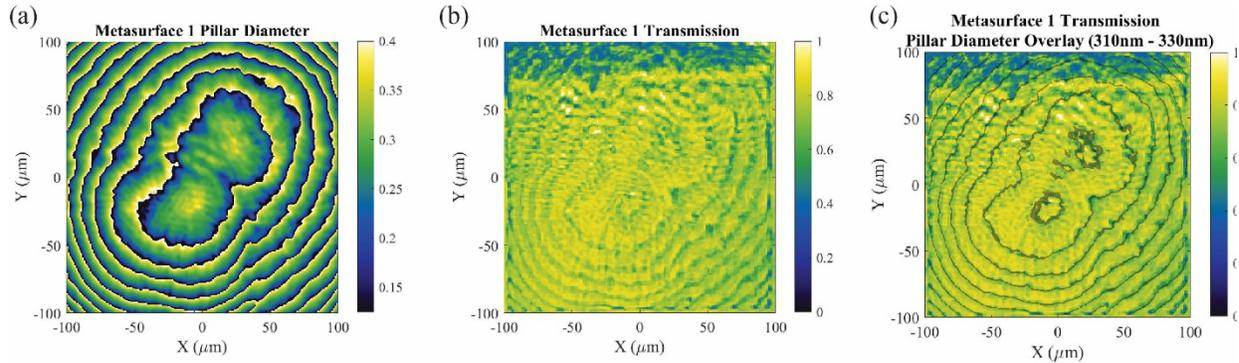

Figure S2: Transmission intensity measurement showing that pillars with diameters of $310nm - 330nm$ give lower transmission. **(a)** The pillar diameters of the nanopillar array implementing metasurface 1 (scale bar in micron). **(b)** Measured transmission intensity of the fabricated metasurface. **(c)** Contours of nanopillars with 310nm-330nm diameters (dark gray bands) overlaid on the measured transmission intensity. These contours match the observe low-transmission contours.

These low-transmission contours result in phase and amplitude errors in the output field distribution of the meta-optic. The hologram image quality is degraded as a result. This is illustrated in Figure S3, where the effects of the low-transmission contours from each metasurface reduce the quality of the desired output field of the meta-optic. The desired output intensity distribution is shown in Figure S3(a), compared to the measured output intensity in Figure S3(b). The measured intensity exhibits low-intensity contours like the low-transmission contours of metasurface 1. Therefore, the errors of the transmitted field introduced by the metasurface directly result in output intensity distribution errors. Even though only intensity errors were measured in the output field distribution, it is expected that phase errors are also present.

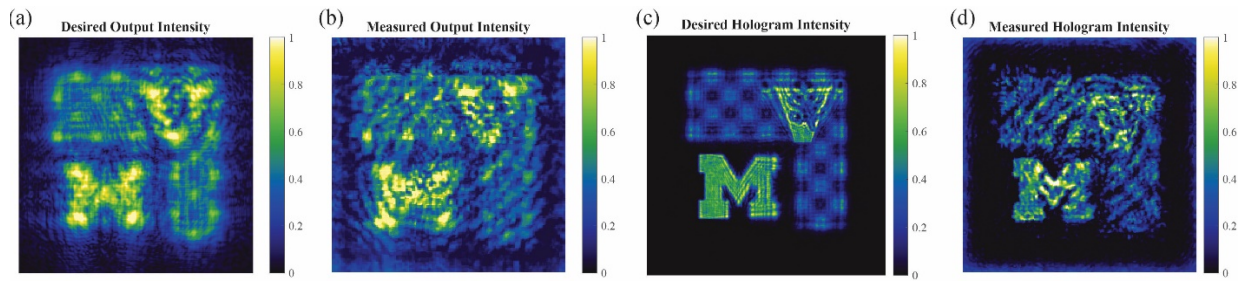

Figure S3: A meta-optic was designed to form a solid-image three-dimensional hologram from a uniform illumination, where the metasurface contained nanopillars with diameters in the entire 125nm-400nm range. The low-transmission contours due to nanopillars with diameters of 310nm-330nm introduce errors into the output field distribution of the meta-optic and degrade the quality of the hologram. Intensity distributions are normalized to their maximums. **(a)** The desired output intensity of the meta-optic. **(b)** The measured output intensity of the meta-optic. Note the low-transmission contours are like those in Figure S2(b). **(c)** The desired intensity of the hologram at a plane where the left portion of the M is in focus. **(d)** Measured intensity of the hologram. The low-transmission contours are visible in the hologram image, degrading its quality.

Due to the amplitude and phase errors in the output field profile, the hologram image is degraded in quality. This is seen when comparing the desired hologram image in Figure S3(c) to the measured image in Figure S3(d). While the general image components are recognizable in the measured image, the low-intensity contours are visible and clearly reduce image quality.

With these results in mind, the three-dimensional hologram was re-designed and fabricated without using nanopillar diameters in the range of 310nm-330nm. The low-transmission contours seen in Figure S2 were mitigated. The measured results are shown in Figure 4(d) of the main text.

These low-transmission contours occur for the beam-splitter and former examples but are not as impactful since this function is measured in the far-field. Introducing output field errors over thin contours creates noise in the plane wave spectrum of the output field distribution. However, as this output field propagates to the far-field, the beam profiles separate from the noise to match the desired profiles. Power is lost to this spectral noise, reducing the efficiency. The measured aperture intensity distribution exhibits thin contours of amplitude errors (see Figure 3 of the main text), but the far-field beam profiles match the expected result.

## Supplementary Note 2: Alignment Hologram

Optimal performance of compound meta-optics requires accurate alignment of the constitutive metasurfaces. A significant phase error is introduced into the output field when the metasurfaces are misaligned. This phase error alters wave propagation so that the desired function is degraded or no longer achieved. When forming a hologram, increasing the phase error on the output field will distort the hologram image.

We utilize an alignment hologram approach to accurately align the metasurfaces. Nanopillar arrays are placed near the two metasurfaces forming the meta-optic. The array near metasurface 1 was designed using the phase-retrieval algorithm to form a cross-shaped intensity pattern at the desired distance of metasurface 2. As a result, the hologram transfers location information from layer 1 to the same plane as layer 2. The nanopillar array near metasurface 2 contains a pattern of low transmission nanopillars tracing the alignment hologram outline. By positioning layer 2 so that the low-transmission outline is not visible when illuminated by the alignment hologram, the two metasurfaces are accurately aligned.

The nanopillar array forming the alignment hologram is $100\mu m \times 100\mu m$ in dimension. Figure S4(a) shows the phase shift profile of this array, which forms the intensity hologram in Figure S4(b). The low-transmission outline is aligned with this hologram by adjusting the position of layer 2, as shown in Figure S4(c). This process accurately aligns the two metasurfaces. Small alignment improvements are made by observing the intensity at a distance from the meta-optic output. As the alignment is improved, the phase error on the output field decreases, leading to an intensity image which matches the desired results.

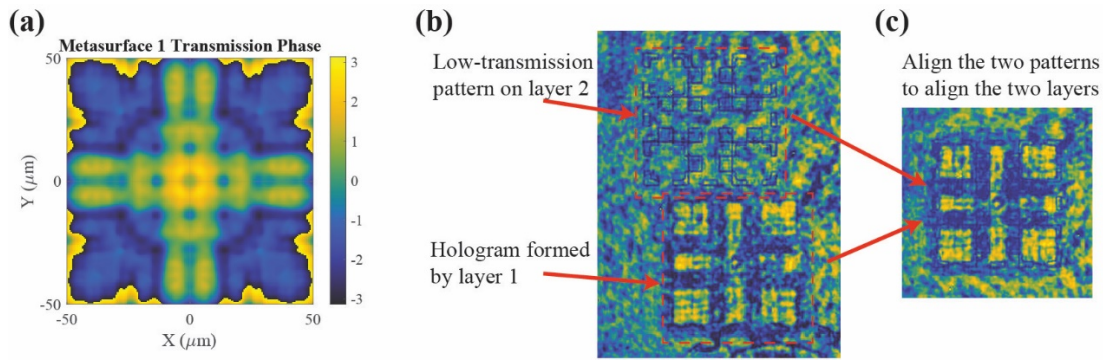

Figure S4: Design and performance of the alignment hologram used to align the two metasurfaces. **(a)** The phase shift profile of a metasurface which forms the desired hologram at the distance where metasurface 2 should be located. **(b)** Intensity observed at the plane of the second metasurface. The low-transmission pattern is introduced at the same plane as metasurface 2, while the alignment hologram comes into focus at the desired separation distance from metasurface 1. **(c)** Aligning the low-transmission pattern with the alignment hologram accurately aligns the two metasurfaces with each other.

## Supplementary Note 3: Meta-optic Designs

The design of each meta-optic example and full-wave simulation results for each metasurface are given in this section. A finite difference time domain (FDTD) simulation is performed for each metasurface using MEEP[7] (an open-source FDTD solver). The simulated performances of each metasurface are combined[8] using spectral domain methods to obtain the simulated output field of the meta-optic device.

## S3.1.  Combined Beam-forming and Splitting: Two Beams

The first meta-optic example reshapes a circular, uniform illumination to form the interference pattern between two Gaussian beams at the output. Specifically, the output field profile is the direct sum of the two Gaussian beam profiles.

$$E_{out} = -e^{-(r/45.5\lambda_0)^2}e^{ik_0x\sin(-2°)} + \sqrt{0.5}e^{-(r/32.5\lambda_0)^2}e^{ik_0x\sin(2°)} \tag{S1}$$

The amplitude of the desired output field distribution is shown in Figure S5(a) and the phase in Figure S5(b). The meta-optic design process was used to determine the transmission phase profiles of each metasurface so the desired field is formed at the output of the device. The resulting transmission phase profiles are shown in Figure S5(c) for metasurface 1 and Figure S5(d) for metasurface 2. These transmission phase profiles are sampled at the unit cell periodicity and translated to a distribution of pillar diameters.

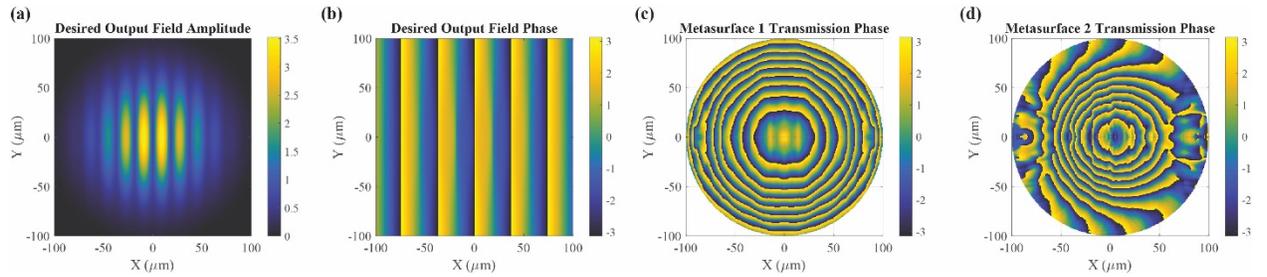

Figure S5: Design of the combined beam-forming and splitting meta-optic with two beams. **(a, b)** The amplitude and phase distributions of the desired output field, which is the interference pattern between two Gaussian beams. **(c, d)** The transmission phase distributions implemented by each metasurface to form the desired field distribution from a circular, uniform illumination.

The complete meta-optic design was simulated to verify the device performance. Metasurface 1 was simulated using the field profile of Figure S6(a) as the incident source field. The transmitted field distribution was recorded at a distance of $0.4\lambda_0$ from the metasurface and is shown in Figure S6(b). A phase shift profile has been imparted on the incident field, and high transmission is maintained. This field profile was numerically propagated across the separation distance to the plane containing the second metasurface and is shown in Figure S6(c). This field distribution was used as the source field of the simulation of metasurface 2. The transmitted field distribution of this simulation is shown in Figure S6(d), where a transmission phase shift occurs while maintaining high transmission.

The simulated output field matches the desired output field, showing that the meta-optic design accurately implements the desired complex-valued field conversion from input to output without relying on loss. The simulated output electric field accuracy is demonstrated by calculating the error in the amplitude and phase distributions relative to the desired output field.  Figure S7 displays the amplitude and phase error profiles, showing a low error across the meta-optic output.

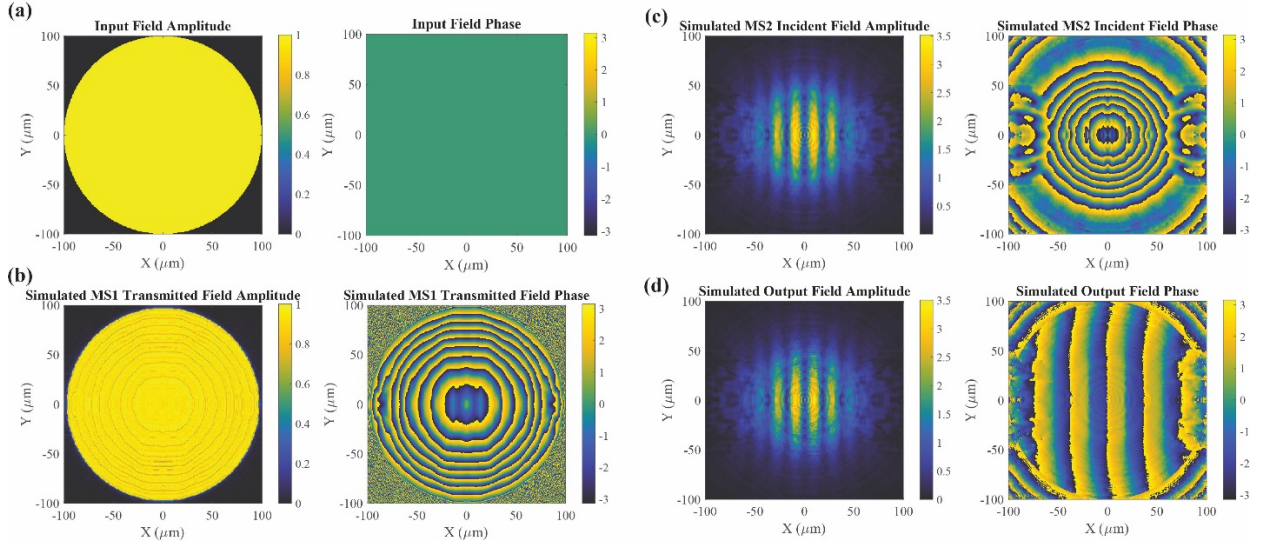

Figure S6: FDTD simulation results of the beam-former and splitter meta-optic example with two beams. **(a)** The field distribution used as the input to the meta-optic is a circular, uniform illumination. **(b)** The simulated transmitted field from metasurface 1 shows that the metasurface applies the desired transmission phase shift with high transmission magnitude. **(c)** The transmitted field from metasurface 1 is numerically propagated to the plane containing metasurface 2. At this plane, the field amplitude distribution matches the desired field amplitude, but the phase is incorrect. **(d)** The simulated transmitted field from metasurface 2 shows that the metasurface applies the desired transmission phase with high transmission magnitude. This field distribution accurately matches the desired output field, indicating that the meta-optic should perform as expected.

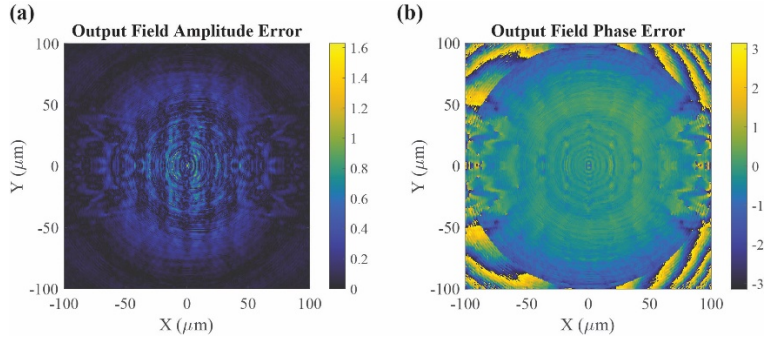

Figure S7: The amplitude and phase error of the simulated output field relative to the desired output field is low for the two-beam former and splitter example, showing the accuracy of the meta-optic performance. **(a)** The amplitude error is shown as the root-mean-square error. **(b)** The phase error of the simulated output field.

## S3.2.  Combined Beam-forming and Splitting: Three Beams

The second beam-former and splitter example forms three output Gaussian beams with different peak intensities and propagation directions. The desired output field is calculated as the sum of three beams

$$E_{out} = e^{-(r/45\lambda_0)^2} \big[ e^{ik_0[x\sin(2°)+y\sin(0°)]} + \sqrt{0.75}\, e^{ik_0[x\sin(-1°)+y\sin(3°)]} \\ + \sqrt{0.5}\, e^{ik_0[x\sin(-2°)+y\sin(-2°)]} \big]. \tag{S2}$$

The amplitude of the desired output field is shown in Figure S8(a) and the phase in Figure S8(b). The transmission phase profiles of the metasurfaces needed to form the desired output field distribution are given in Figure S8(c, d).

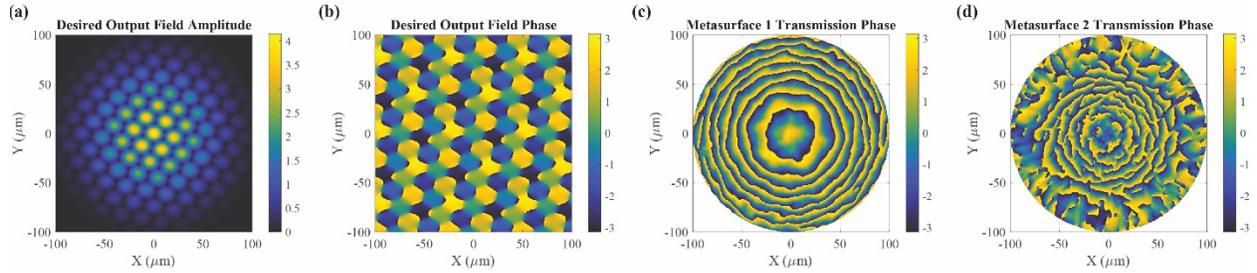

Figure S8: Design of the combined beam-forming and splitting meta-optic with three beams. **(a, b)** The amplitude and phase distributions of the desired output field, which is the interference pattern between three Gaussian beams. **(c, d)** The transmission phase distributions implemented by each metasurface to form the desired output field from the uniform, circular illumination.

The meta-optic simulation process was performed, with results displayed in Figure S9. Comparing the simulated output field in Figure S9(d) to the desired output field in Figure S8(a, b), we see that the meta-optic design accurately performs the desired complex-valued field conversion. This is further verified by Figure S10 which shows the amplitude and phase error distribution of the simulated output field.

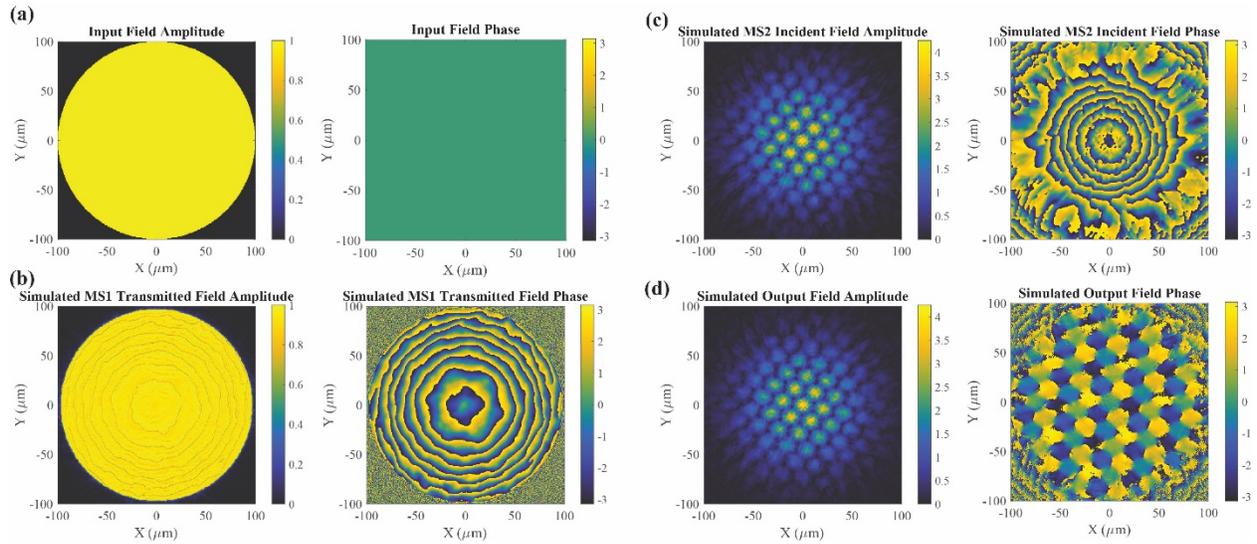

Figure S9: FDTD simulation results of the beam-former and splitter meta-optic example with three beams. **(a)** The input field distribution to the meta-optic is a circular, uniform distribution. **(b)** The simulated transmitted field from metasurface 1 exhibits the desired phase shift with a high transmission magnitude. **(c)** The transmitted field from metasurface 1 is numerically propagated to the plane containing metasurface 2. At this plane, the desired amplitude distribution is formed, but the phase is incorrect. **(d)** The simulated transmitted field from metasurface 2. This field accurately matches the desired output field distribution, indicating that the meta-optic should perform as expected.

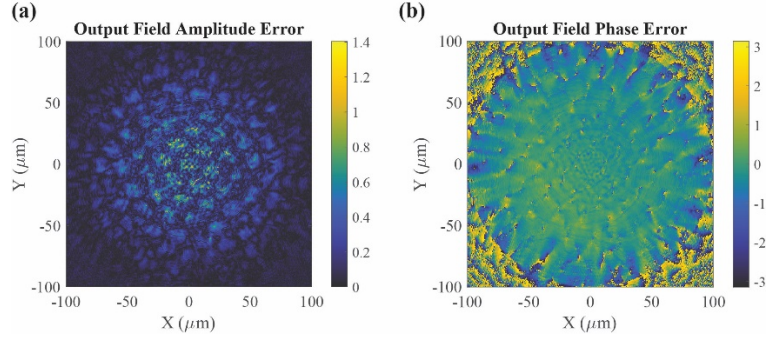

Figure S10: The amplitude and phase distribution error of the simulated output field relative to the desired output field is low for the three-beam former and splitter example, showing the accuracy of the meta-optic performance. **(a)** The amplitude error is shown as the root-mean-square error. **(b)** The phase error of the simulated output field.

### S3.3. Point-source Three-dimensional Hologram

The first three-dimensional hologram example forms a constellation of point sources. The point sources trace the outline of the University of Michigan and Vanderbilt University logos, tilted in space to provide depth. Each logo was tilted 15 degrees about its center. The meta-optic output field was constructed by propagating the field from each point source to the meta-optic output plane and summing them together. The complex conjugate of this field profile was used as the desired output field distribution, so the hologram was formed beyond the meta-optic (the observer side of the output plane). The operating wavelength of this example is $\lambda_0 = 1.1 \mu m$, with the unit cell transmission characteristics shown in Figure S11.

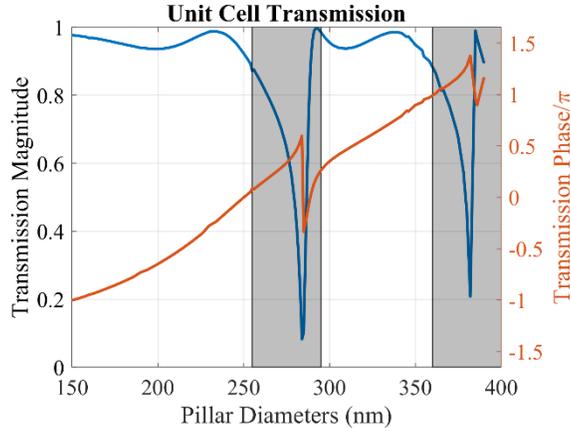

Figure S11: Unit cell transmission characteristics for the $\lambda_0 = 1.1 \mu m$ implementation of the point-source hologram meta-optic. For this unit cell design, the period is 400nm, the pillar height 750nm, and the pillar diameter ranges from 150nm-360nm while skipping the 255nm-290nm range due to low transmission.

The amplitude and phase distributions of the meta-optic output field forming the point source hologram are shown in Figure S12(a, b). The metasurface transmission phase profiles are shown in Figure S12(c, d).

The meta-optic was simulated, with results given in Figure S13. Comparing the simulated output field in Figure S13(d) to the desired output field in Figure S12(a, b), we see that the meta-optic design accurately performs the desired complex-valued field conversion. This is further verified in Figure S14 when observing the amplitude and phase error distribution of the simulated output field.

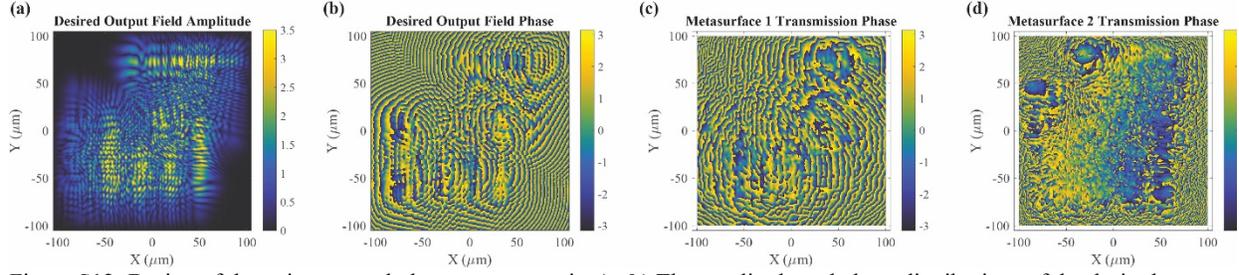

Figure S12: Design of the point source hologram meta-optic. **(a, b)** The amplitude and phase distributions of the desired output field, which is the complex conjugate of the interference pattern formed by the point sources in the hologram. **(c, d)** The transmission phase shift profiles implemented by each metasurface to form the desired output field from the uniform, square illumination.

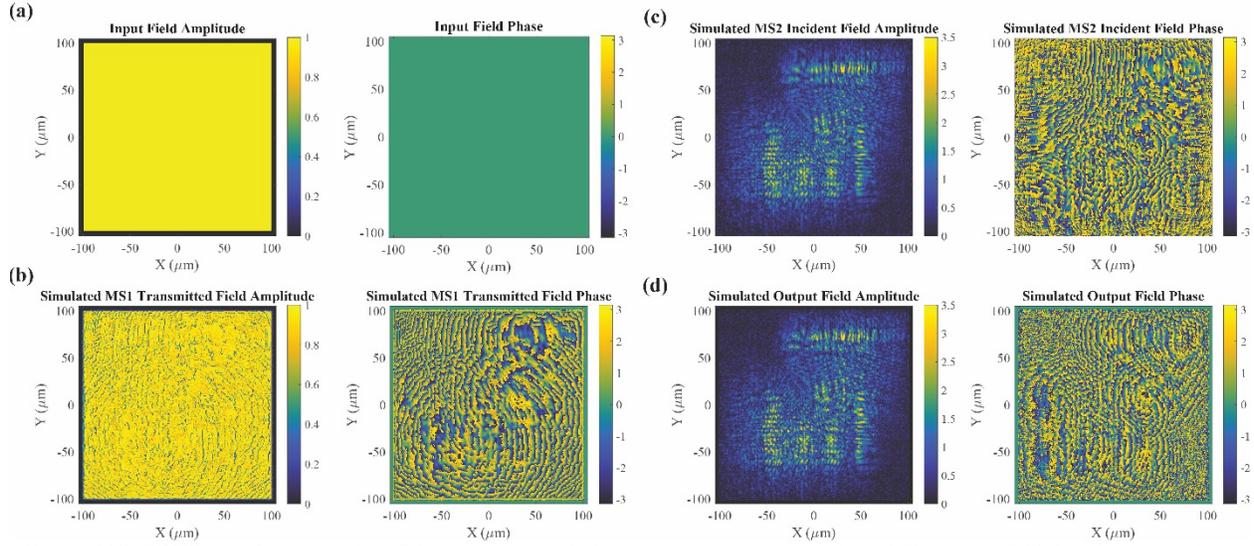

Figure S13: FDTD simulation results of the point source hologram meta-optic example. **(a)** The input field distribution to the meta-optic is a square, uniform distribution. **(b)** The simulated transmitted field from metasurface 1 exhibits the desired phase shift with a high transmission magnitude. **(c)** The transmitted field from metasurface 1 is numerically propagated to the plane containing metasurface 2. At this plane, the desired amplitude distribution is formed, but the phase is incorrect. **(d)** The simulated transmitted field from metasurface 2. This field accurately matches the desired output field distribution, indicating that the meta-optic should perform as expected.

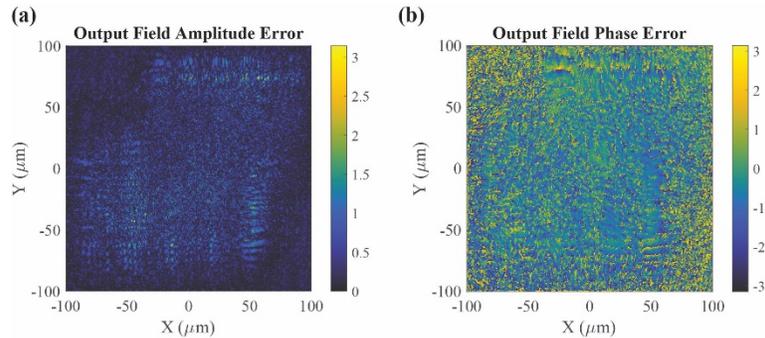

Figure S14: The amplitude and phase distribution error of the simulated output field relative to the desired output field is low for the point-source hologram example, showing the accuracy of the meta-optic performance. **(a)** The amplitude error is shown as the root-mean-square error. **(b)** The phase error of the simulated output field.

## S3.4.  Solid-image Three-dimensional Hologram

The second three-dimensional hologram example forms a hologram consisting of solid images tilted in space to provide depth to the scene. The hologram was constructed using computer-generated hologram methods[9-11]. Each image component was converted to an equivalent source of light, rotated in space to appear at an angle to the observer, and assembled to form the overall scene[8]. This hologram is composed of four separate image components: i) spiral, ii) series of stars, iii) University of Michigan logo, and iv) Vanderbilt University logo. The image components are rotated around an axis and placed at specific coordinates to assemble the 3D scene. The axis of rotation, rotation angle, and center position for each component are given in Table 1. While this 3D scene is relatively simple, the process can be followed to form a scene with many different image components arranged to form three-dimensional shapes of more complex objects.

Table 1: Manipulation parameters of each image component to assemble the three-dimensional hologram scene.

| Image Component | Rotation Axis | Rotation Angle (degrees) | Center Position |
|---|---|---|---|
| Spiral | $y = x$ | 20 | $(0, 0, 104\mu m)$ |
| Stars | No rotation | No rotation | $(0, 0, 104\mu m)$ |
| University of Michigan logo | $y$ | 25 | $(-35\mu m, -35\mu m, 156\mu m)$ |
| Vanderbilt University logo | $x$ | 25 | $(55\mu m, 50\mu m, 104\mu m)$ |

The field distribution forming these hologram image components is calculated at the meta-optic output plane (located at $z = 0$) and is shown in Figure S15(a, b) as the amplitude and phase distributions. The metasurface transmission phase profiles needed to perform the complex-valued field conversion are shown in Figure S15(c, d).

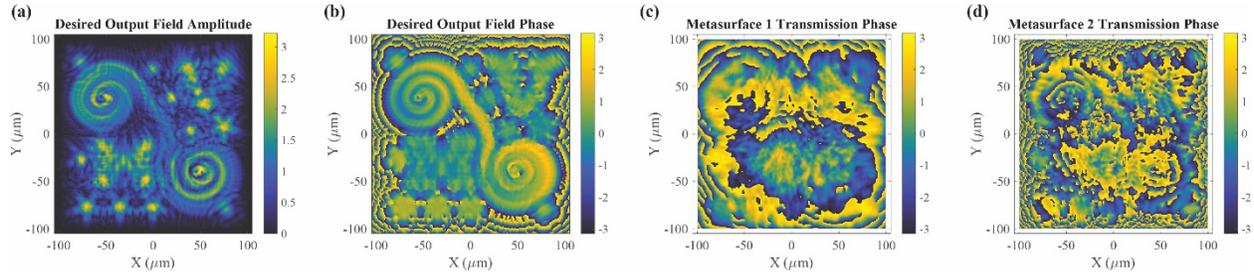

Figure S15: Design of the solid-image 3D hologram. **(a, b)** The amplitude and phase distributions of the desired output field, which is constructed using computer-generated hologram methods. **(c, d)** The transmission phase distributions implemented by each metasurface to form the desired output field distribution from a square, uniform illumination.

The meta-optic was simulated, with results given in Figure S16. By comparing the simulated output field in Figure S16(d) to the desired output field in Figure S15(a, b), we see that the meta-optic design accurately performs the desired complex-valued field conversion. This is verified in Figure S17 when observing the amplitude and phase error distribution of the simulated output field.

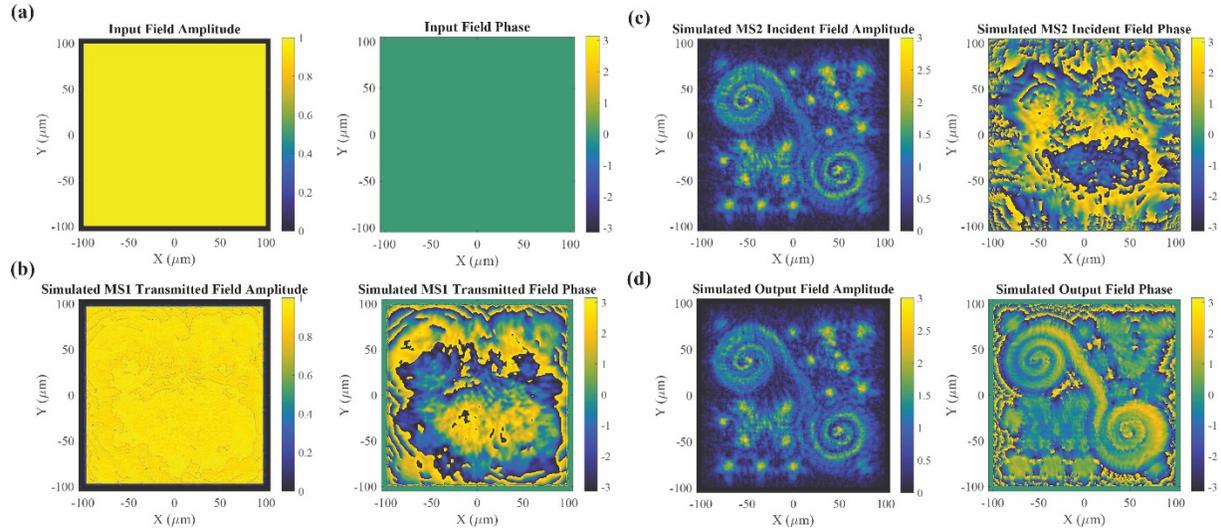

Figure S16: FDTD Simulation results of the solid-image 3D hologram example. **(a)** The input field distribution to the meta-optic is a square, uniform distribution. **(b)** The simulated transmitted field from metasurface 1 exhibits the desired phase shift with a high transmission magnitude. **(c)** The transmitted field from metasurface 1 is numerically propagated to the plane containing metasurface 2. At this plane, the desired amplitude distribution is formed, but the phase is incorrect. **(d)** The simulated transmitted field from metasurface 2. This field accurately matches the desired output field distribution.

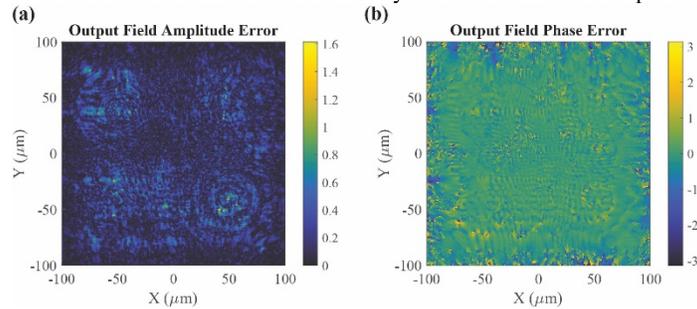

Figure S17: The amplitude and phase error of the simulated output field relative to the desired output field is low for the solid-image 3D hologram example, showing the accuracy of the meta-optic performance. **(a)** The amplitude error is shown as the root-mean-square error. **(b)** The phase error of the simulated output field.

## Supplementary Note 4: Three-Dimensional Hologram — Comparison of Complex-Valued to Phase-only Design

Here, we compare phase and amplitude control provided by the meta-optic to the phase-only control available from a single metasurface. Phase-only control is achieved by an individual metasurface applying a desired phase profile to the incident uniform illumination. The resulting transmitted field exhibits the desired phase profile, but an incorrect amplitude profile. An in-depth comparison of phase-only to complex-valued phase and amplitude control is given by A. Overvig et. al[12].

In the case of the solid-image three-dimensional hologram, the phase-only metasurface was designed to implement the phase distribution of the desired output field profile, shown in Figure S15(b). As the transmitted field propagates away from the metasurface, it forms a version of the desired hologram image. Figure S18 shows the hologram intensity at different depths for the cases of simulated and measured phase-only control compared to measured phase and amplitude control. From this comparison, we see that the measured holograms exhibit higher image quality when both the phase and amplitude profiles of the field distribution are manipulated to match the desired distribution.

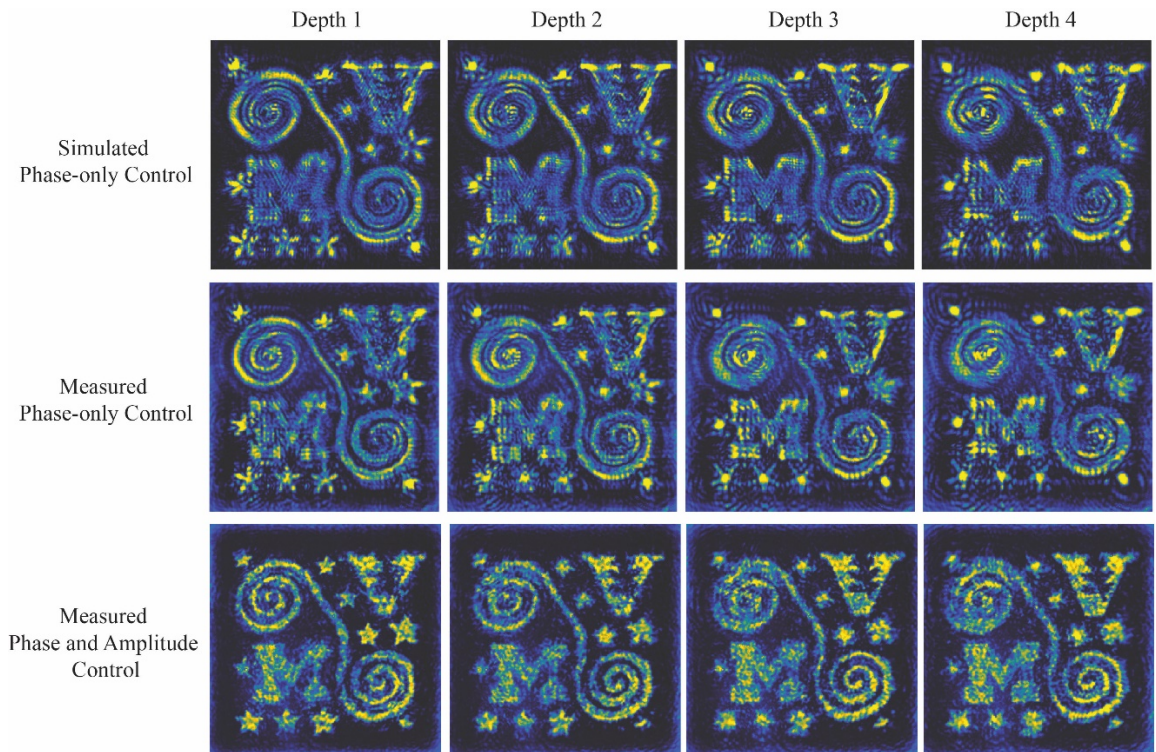

|  | Depth 1 | Depth 2 | Depth 3 | Depth 4 |
|--|---------|---------|---------|---------|
| Simulated Phase-only Control | | | | |
| Measured Phase-only Control | | | | |
| Measured Phase and Amplitude Control | | | | |

Figure S18: Comparison of hologram images for simulated and measured phase-only control to the measured phase and amplitude control approach. Using both phase and amplitude control results in hologram images that have improved image quality throughout the volume of the three-dimensional hologram.